\documentclass[11pt,nofootinbib,showpacs]{revtex4}
\usepackage{amsfonts,amssymb,amsmath,graphicx,multirow}
\def\dac{\displaystyle\frac}

\def\[{\left[}
\def\]{\right]}
\def\({\left(}
\def\){\right)}

\newcommand{\const}{\mathop{\rm const}\nolimits}

\begin{document}

\baselineskip7mm

\title{Exponential cosmological solutions in Einstein-Gauss-Bonnet gravity with two subspaces: general approach}

\author{Sergey A. Pavluchenko}
\affiliation{Programa de P\'os-Gradua\noexpand\c c\~ao em F\'isica, Universidade Federal do Maranh\~ao (UFMA), 65085-580, S\~ao Lu\'is, Maranh\~ao, Brazil}
\affiliation{A. Alikhanyan National Science Laboratory  (Yerevan Physics Institute), Alikhanyan Brothers st. 2, 0036, Yerevan, Armenia}

\begin{abstract}
In this paper we perform systematic investigation of all possible exponential solutions in Einstein-Gauss-Bonnet gravity with the spatial section being a product of two subspaces. We describe a
scheme which always allow to find solution for a given $\{p, q\} > 2$ (number of dimensions of two subspaces) and $\zeta$ (ratio of the expansion rates of these two subspaces). Depending on
the parameters, for given $\{\alpha, \Lambda\}$ (Gauss-Bonnet coupling and cosmological constant)
there could be up to four distinct solutions (with different $\zeta$'s). Stability requirement introduces relation between $\zeta$, $\{p, q\}$ and sign of the expansion rate. Nevertheless,
for any $\{p, q\} > 2$ we can always choose sign for expansion rates so that the resulting solution would be stable. The scheme for finding solutions is described and the bounds on the parameters
are drawn. Specific cases with $\{p, q\} = \{1, 2\}$ are also considered. Finally, we separately described physically sensible case with one of the subspaces being three-dimensional and expanding
(resembling our Universe) while another to be contracting (resembling extra dimensions), describing successful compactification; for this case we also  drawn bounds on the parameters where such regime
occurs.
\end{abstract}







\maketitle

\section{Introduction}

It is not widely known, but the idea of extra dimensions\footnote{By ``extra dimensions'' we mean metric setup with number of spatial dimensions exceeds ``standard'' three.} precedes General Relativity
(GR)---indeed, first known extra-dimensional theory was proposed by Nordstr\"om in 1914~\cite{Nord1914} and it unified Nordstr\"om's second gravity theory~\cite{Nord_2grav} and Maxwell's electromagnetism.
Soon after Einstein proposed GR~\cite{einst} which confronted Nordstr\"om's gravity, and in 1919 confrontation ended in favor of GR: Nordstr\"om's gravity, like most of scalar gravity theories,
predicts no bending of light near massive bodies, yet, observations made during the Solar Eclipse in 1919 clearly demonstrated that the deflection angle is in agreement with GR predictions.

So that the Nordstr\"om's gravity was abandoned but not the idea---soon Kaluza proposed~\cite{KK1} very similar model but based on GR: within his theory five-dimensional Einstein equations could be
decomposed into four-dimensional Einstein equations
plus Maxwell's electromagnetism. To perform this decomposition, the extra dimensions should be ``curled'' or compactified into a circle of a very small size and ``cylindrical conditions'' should be imposed.
Later Klein proposed~\cite{KK23} a nice quantum mechanical interpretation of this extra dimension, so that the resulting theory, named Kaluza-Klein after its founders, was formally formulated.
Remarkably, this theory unified all
known interactions at that time. With time, more interactions were known and it became clear that to unify them all, more extra dimensions are needed. Nowadays, one of the promising theories to unify
all interactions is M/string theory.

One of the distinguishing features of M/string theories is the presence of the curvature-squared corrections in the Lagrangian of their gravitational counterpart. First it was noticed by
Scherk and Schwarz~\cite{sch-sch}---they demonstrated the presence of the $R^2$ and $R_{\mu \nu} R^{\mu \nu}$ terms in the Lagrangian of the Virasoro-Shapiro model~\cite{VSh}.
Candelas et al.~\cite{Candelas_etal} found presence of curvature-squared term of the $R^{\mu \nu \lambda \rho} R_{\mu \nu \lambda \rho}$ type in the low-energy limit
of the $E_8 \times E_8$ heterotic superstring theory~\cite{Gross_etal} to match the kinetic term of the Yang-Mills field. Later on it was demonstrated by Zwiebach~\cite{zwiebach}
that the only combination of quadratic terms which leads to a ghost-free nontrivial gravitation interaction is the Gauss-Bonnet (GB) term:

$$
L_{GB} = R_{\mu \nu \lambda \rho} R^{\mu \nu \lambda \rho} - 4 R_{\mu \nu} R^{\mu \nu} + R^2.
$$

\noindent This term, first found by Lanczos~\cite{Lanczos} (and so sometimes it is referred to as the Lanczos term) is an Euler topological invariant in (3+1)-dimensional
space-time, but starting from (4+1) and in higher dimensions it gives nontrivial contribution to the equations of motion.
Zumino~\cite{zumino} extended Zwiebach's result on higher-than-squared curvature terms, supporting the idea that the low-energy limit of the unified
theory might have a Lagrangian density as a sum of contributions of different powers of curvature. In this regard the Einstein-Gauss-Bonnet (EGB) gravity could be seen as a subcase of more general Lovelock
gravity~\cite{Lovelock}, but in the current paper we restrain ourselves with only quadratic corrections and so to the EGB case.

Both our everyday experience and the experiments in particle and space physics clearly demonstrate that there are only three spatial dimensions (for example, for Newtonian gravity
defined in more than three spatial dimensions there are no stable orbits while we are rotating around the Sun for ages). So that to bring together the extra-dimensional theories and
the experiment, we need to explain where are these extra dimensions. The commonly accepted answer is that the extra dimensions are compactified within a very small scale---similar
to the ``curling'' of extra dimension in the original Kaluza-Klein paper. But this answer, in its turn, gives rise to another question---how comes that they became compact?
The answer to this question is not simple. First attempts to answer this question involve solution known as ``spontaneous compactification''~\cite{add_1, Deruelle2}.
Similar solutions but more relevant to cosmology were proposed in~\cite{add_4} (see also~\cite{prd09}). More natural way to achieve compactified extra dimensions
is ``dynamical compactification'' (roughly speaking, in ``spontaneous compactification'' scenario you put extra dimensions compact ``by hand'' and see if the resulting setup is viable within the gravity
theory under consideration; in contrast, in ``dynamical compactification'' scenario you do not impose small extra dimensions in the beginning but they become small in the process of the evolution).
The works in this direction involves different approaches~\cite{add_8} and different setups~\cite{MO}. Also, apart from the
cosmology, the studies of extra dimensions involve investigation of properties of black holes in Gauss-Bonnet~\cite{BHGB} and
Lovelock~\cite{BHL} gravities, features of gravitational collapse in these
theories~\cite{coll}, general features of spherical-symmetric solutions~\cite{addn_8}, and many others.

The equations of motion for EGB and for more general Lovelock gravity are much more complicated than in GR, and it is very nontrivial to find exact solutions within EGB gravity,
so that one usually apply metric {\it ansatz} of some sort. For cosmology the usual {\it ans\"atzen} are power-law and exponential---the former resemble Friedmann
stage while the latter---accelerated expansion nowadays or inflationary stage in Early Universe. Power-law solution were studied in~\cite{Deruelle1, Deruelle2} and more
 recently in~\cite{mpla09_iv, prd09, prd10, grg10}, so that by now we can say that we have some understanding of their dynamics. One of the first considerations of the
 exponential solutions in the considered theories was done in~\cite{Is86}, the recent works include~\cite{KPT}, separate description of the exponential solutions with variable~\cite{CPT1}
 and constant~\cite{CST2} volume; see also~\cite{PT} for the discussion about the link between existence of power-law and exponential solutions as well as for the discussion
 about the physical branches of these solutions.
 We have also described the general scheme for finding all possible exponential solutions in arbitrary dimensions and with arbitrary
 Lovelock contributions taken into account in~\cite{CPT3}. Deeper investigation revealed that not all of the solutions found in~\cite{CPT3} are stable~\cite{my15};
 see also~\cite{iv16} for more general approach to the stability of exponential solutions in EGB gravity and~\cite{stab_add, stab_add1, stab_add2} for some particular cases.

 The above approach gave us asymptotic power-law and exponential regimes, but not the entire evolution. Without full evolution we cannot decide if the asymptotic solution which we
 found is realistic or not, could it be reached from vast enough initial conditions or is it just some artifact. To answer this question we need to go beyond exponential or
 power-law {\it ans\"atzen} and consider generic form of the scale factor. In this case, though, one would often resort to numerics. In~\cite{CGP1} we considered the
 cosmological model with the spatial part being a product of three- and extra-dimensional spatially curved manifolds and numerically demonstrated the existence of
 the phenomenologically
sensible regime when the curvature of the extra dimensions is negative and the EGB theory does not admit a maximally symmetric solution. In this case both the
three-dimensional Hubble parameter and the extra-dimensional scale factor asymptotically tend to the constant values (so that three-dimensional part expanding exponentially while
the extra dimensions tends to a constant ``size'').
In~\cite{CGP2} the study of this model was continued and it was demonstrated that the described above regime is the only realistic scenario in the considered model.
 Recent analysis of the same model~\cite{CGPT} revealed that, with an additional constraint on couplings, Friedmann-type late-time behavior
could be restored.

Mentioned above investigation was performed numerically, but to find all possible regimes for a particular model, the numerical methods are not the best practice. We have noted that if one
considers the model with spatial section being the product of three- and extra-dimensional spatially-flat subspaces, the equations of motion simplify and become
first-order\footnote{Situation similar to the Friedmann equations---if the spatial curvature $k \ne 0$ then they are second order with respect to the scale
factor $a(t)$ ($\ddot a$ is the highest derivative), but if $k \equiv 0$ then they could be rewritten in terms of the Hubble parameter $H \equiv \dot a/a$ and become first order
($\dot H$ is the highest derivative).}. In this case the dynamics could be analytically described with use of phase portraits and so we performed this analysis. For vacuum EGB
case we have done in~\cite{my16a} and reanalyzed in~\cite{my18a}. The results suggest that in the vacuum model has two physically viable regimes---first of them is the smooth transition from high-energy GB Kasner to low-energy GR Kasner. This regime
exists for $\alpha > 0$ (Gauss-Bonnet coupling) at $D=1,\,2$ (the number of extra dimensions) and for $\alpha < 0$ at $D \geqslant 2$ (so that at $D=2$ it appears for both signs of $\alpha$). Another viable regime is the smooth transition from high-energy GB
Kasner to anisotropic exponential solution with expanding three-dimensional section (``our Universe'') and contracting extra dimensions; this regime occurs only for $\alpha > 0$ at $D \geqslant 2$.
Apart from the EGB, similar analysis but for cubic Lovelock contribution taken into account was done in~\cite{cubL}.

For EGB model with $\Lambda$-term same analysis was performed in~\cite{my16b, my17a} and reanalyzed in~\cite{my18a}. The results suggest that the only realistic scenario
is the transition from high-energy GB Kasner to anisotropic exponential
regime, and it requires $D \geqslant 2$, see~\cite{EGBL, my18a} for exact limits on ($\alpha, \Lambda$) when this regime exist.
The low-energy GR Kasner is
forbidden in the presence of the $\Lambda$-term so the analogous to the vacuum case transition do not occur.

For completeness sake it is worth mentioning that apart from vacuum and $\Lambda$-term models we considered models with perfect fluid as a source: initially we considered them in~\cite{KPT},
some deeper studies of
(4+1)-dimensional Bianchi-I case was done in~\cite{prd10} and deeper investigation of power-law regimes in pure GB gravity in~\cite{grg10}. Systematic study for all $D$ was started in~\cite{my18d} for
low $D$ and is currently continued for high $D$ cases.

In the studies described above we have made two important assumptions---we considered both subspaces to be isotropic and spatially flat.
But what if we lift these conditions? Indeed, the spatial section
as a product of two isotropic spatially-flat subspaces could hardly be called ``natural'', so that we considered the effects of anisotropy and spatial curvature in~\cite{PT2017}. The effect of the initial anisotropy could be seen on the example of vacuum $(4+1)$-dimensional EGB model with Bianchi-I-type metric
(all directions are independent), where the only future asymptote is nonstandard singularity~\cite{prd10}. Our analysis~\cite{PT2017} suggest that the transition from
 GB Kasner regime to anisotropic exponential model (with expanding
three and contracting extra dimensions) is stable with respect to breaking the symmetry within both three- and extra-dimensional subspaces. However, the details of the dynamics in
$D=2$ and $D \geqslant 3$ are different---in the latter case there formally exist anisotropic exponential solutions with ``wrong'' spatial splitting and (again, formally)
all of them are accessible from generic
initial conditions. For example, in $(6+1)$-dimensional space-time there exist anisotropic exponential solutions with $[3+3]$ and $[4+2]$ spatial splittings, and some of the initial
conditions from the vicinity of $E_{3+3}$ actually end up in $E_{4+2}$---exponential solution with four and two isotropic subspaces. In other words, generic initial conditions
could easily end up with ``wrong'' compactification, giving ``wrong'' number of expanding spatial dimensions (see~\cite{PT2017} for details; see also~\cite{CGT-2020} for $(7+1)$-dimensional space-time).

The effect of the spatial curvature on the cosmological dynamics could be dramatic---for example, positive curvature could change inflationary asymptotic~\cite{infl} in standard Friedmann
cosmology.
In EGB gravity the influence of the spatial curvature (also described in~\cite{PT2017})
reveal itself for both signs of the curvature of extra dimensions\footnote{And in this lies the difference with the standard Friedmann cosmology, where negative curvature plays sort of
``repulsive force'' role.} in $D \geqslant 3$---in that case for negative curvature there exist  ``geometric frustration'' regime, described in~\cite{CGP1, CGP2} and further investigated in~\cite{CGPT} while
for the positive curvature, solutions
with stabilized extra dimensions could coexist with maximally symmetric solution, but for a very narrow range of parameters~\cite{our20}.
So to speak, with negative curvature of extra dimensions stabilized
solutions almost always exist but maximally symmetric solution does not coexist with them while for the positive curvature we require fine-tuning for stabilized solutions to exist but when they do,
they coexist with maximally symmetric solution; see also~\cite{CP21} for some more details on the difference between the cases with positive and negative spatial curvature.

In this manuscript we return to the question of existence and stability of the exponential solutions. We consider only solutions with two subspaces as it is the most simple, most wide-spread and
the only relevant solution in low number of spatial dimensions. Despite the fact that physically meaningful only solutions with expanding three dimensions (``our Universe'') and contracting other
dimensions (``extra dimensions''), it is important to understand general dynamics of such system---this, in turn, could shed light on some underlying principles for successful compactification.

The manuscript structured as follows: first we derive the equations of motion, then solve them for the general case and find properties of the solution found. After that we address linear stability
of the solution, and consider special cases which formally do not fall into the general scheme (there are multipliers which nullify expressions for these particular special cases). Then we consider
the case with three-dimensional subspace which could be seen as successful compactification with expanding three (``our Universe'') and contracting extra dimensions.
Finally, we summarize our results, discuss them and draw conclusions.

\section{Equations of motion}

In the case of two subspaces the metric could be written as

\begin{equation}
\begin{array}{l}
ds^2 = -dt^2 + a(t)^2 d\Sigma_p^2 + b(t)^2 d\Sigma_q^2.
\end{array} \label{metric2}
\end{equation}

\noindent where $d\Sigma_p^2$ and $d\Sigma_q^2$ are spatially-flat line elements corresponding to $p$- and $q$-dimensional submanifolds and we assume speed of light $c \equiv 1$.

General Einstein-Gauss-Bonnet action in vielbein formalism takes a form

\begin{equation}
\begin{array}{l}
S = \int \varepsilon_{A_1 \dots A_N} \( \tilde\Lambda e^{A_1} \dots e^{A_N} + \tilde G_{N, eff} R^{A_1 A_2} e^{A_3} \dots e^{A_N} + \tilde\alpha R^{A_1 A_2} R^{A_3 A_4} e^{A_5} \dots e^{A_N}  \),
\end{array} \label{viel1}
\end{equation}

\noindent where $N$ is the total number of spacetime dimensions (so that $N=1+p+q$ for our particular case), $A_i$ is the index which count them, $\tilde\Lambda$ is the boundary term (generally it is
not equal to the $\Lambda$-term, see e.g.~\cite{CGP1, CGP2}), $\tilde G_{N, eff}$ is effective Newtonian constant and it is multiplied by Einstein-Hilbert contribution and finally
$\tilde\alpha$ is Gauss-Bonnet coupling and it is multiplied by Gauss-Bonnet term. In our metric {\it ans\"atz} components of the curvature two-form take form

\begin{equation}
\begin{array}{l}
R^{0\alpha} = \dac{\ddot a(t)}{a(t)} e^0 \wedge e^\alpha,~~ R^{0a} = \dac{\ddot b(t)}{b(t)} e^0 \wedge e^a,~~
R^{\alpha\beta} = \dac{\dot a(t)^2}{a(t)^2} e^\alpha \wedge e^\beta, \\ R^{ab} = \dac{\dot b(t)^2}{b(t)^2} e^a \wedge e^b,~~
R^{\alpha a} = \dac{\dot a(t)}{a(t)} \dac{\dot b(t)}{b(t)}e^\alpha \wedge e^b.
\end{array} \label{viel2}
\end{equation}

In this paper we are interested in exponential solutions, so that we substitute $a(t)=\exp (Ht)$, $b(t)=\exp(ht)$ and obtain equations of motion via standard procedure (see e.g.~\cite{prd09, CPT3}):

\begin{equation}
\begin{array}{l}
p(p-1)H^2 + 2Hhpq + q(q-1)h^2 + \alpha\[ p(p-1)(p-2)(p-3)H^4 + 4pq(p-1)(p-2)H^3h + \right. \\ + \left. 6pq(p-1)(q-1)H^2h^2 + 4pq(q-1)(q-2)Hh^3 + q(q-1)(q-2)(q-3)h^4 \] = \Lambda, \\
p(p-1)H^2 + 2q(p-1)Hh + q(q+1)h^2 + \alpha\[ p(p-1)(p-2)(p-3)H^4 + 4q(p-1)^2(p-2)H^3h + \right. \\ + \left. 2q(p-1)(3pq-p-4q)H^2h^2 + 4q^2(q-1)(p-1)Hh^3 + (q+1)q(q-1)(q-2)h^4  \] = \Lambda, \\
p(p+1)H^2 + 2p(q-1)Hh + q(q-1)h^2 + \alpha \[  (p+1)p(p-1)(p-2)H^4 + 4p^2(p-1)(q-1)H^3h + \right. \\ + \left. 2p(q-1)(3pq-4p-q)H^2h^2 + 4p(q-1)^2(q-2)Hh^3 + q(q-1)(q-2)(q-3)h^4  \] = \Lambda.
\end{array} \label{eom2}
\end{equation}

One can notice ``unusual'' multiplier in from of $H^2 h^2$ terms ($(3pq-p-4q)$ and $(3pq-4p-q)$), as this term originate from three sorts of curvature combinations: $R^{0\alpha} R^{ab}$,
$R^{0a} R^{\alpha\beta}$ and $R^{a\alpha} R^{b\beta}$. We also renormalized $\tilde\Lambda$, $\tilde G_{N, eff}$ and $\tilde\alpha$ to take convenient form with $\tilde G_{N, eff} \to 1$.

\section{Solutions and their properties}
\label{exist}

One can see that the resulting equations of motion are fourth power with respect to $H$ and $h$, making it quite hard to analyze ``head on''. To simplify the system, we introduce new variables: $H/h=\zeta$,
$\alpha\Lambda=\xi$ and $h^2=\theta/\alpha$; in new variables the system takes form which is very similar to (\ref{eom2}), but for Einstein-Hilbert contribution we have $\theta$ multiplier while for
Gauss-Bonnet it is $\theta^2$, other variables transform as $\Lambda\to\xi$, $H^4\to\zeta^4$, $H^3h\to\zeta^3$, $H^2\to\zeta^2$, $H^2h^2\to\zeta^2$, $Hh^3\to\zeta$, $Hh\to\zeta$, $h^2\to 1$, $h^4\to 1$.
Despite the fact that the equations looks quite similar, we write down all of them, as we are going to describe in detail the derivation, and for this reason it is useful to have all actual equations:

\begin{equation}
\begin{array}{l}
\theta \[p(p-1)\zeta^2 + 2\zeta pq + q(q-1) \] + \theta^2\[ p(p-1)(p-2)(p-3)\zeta^4 + 4pq(p-1)(p-2)\zeta^3 + \right. \\ + \left. 6pq(p-1)(q-1)\zeta^2 + 4pq(q-1)(q-2)\zeta + q(q-1)(q-2)(q-3) \] = \xi; \\
\theta \[p(p-1)\zeta^2 + 2q(p-1)\zeta + q(q+1) \] + \theta^2\[ p(p-1)(p-2)(p-3)\zeta^4 + 4q(p-1)^2(p-2)\zeta^3 + \right. \\ + \left. 2q(p-1)(3pq-p-4q)\zeta^2 + 4q^2(q-1)(p-1)\zeta + (q+1)q(q-1)(q-2)
 \] = \xi, \\
\theta \[ p(p+1)\zeta^2 + 2p(q-1)\zeta + q(q-1)\] + \theta^2 \[  (p+1)p(p-1)(p-2)\zeta^4 + 4p^2(p-1)(q-1)\zeta^3 + \right. \\ + \left. 2p(q-1)(3pq-4p-q)\zeta^2 + 4p(q-1)^2(q-2)\zeta + q(q-1)(q-2)(q-3)
 \] = \xi.
\end{array} \label{eom2_2}
\end{equation}

 The equations remain fourth order, but we will not solve them for $\zeta$. Instead, we will use $\zeta$ as a parameter for the
following reason: as we mentioned, ultimately we are interested in finding stable successful compactification regimes, so that instead of asking if there is compactification for a given
$\{\alpha, \Lambda\}$ in given number of extra dimensions, we ask for which $\{\alpha, \Lambda\}$ in given number of extra dimensions compactification occurs. Indeed, if we want for extra dimensions to
contract, we choose specific $\zeta < 0$ and solve the system to obtain $\theta$ and $\xi$.

The system (\ref{eom2_2}) has three equations, but, similar to the ``standard'' Friedmann cosmology, only two of them are independent, and that is why there are only two independent variables:
$\theta$ and $\xi$. We express $\xi$ from the constraint equation (first of (\ref{eom2_2}))
and substitute it to remaining two equations. Equivalently this could be seen as subtracting constraint equation from the remaining dynamical equations.
One can note that some of the terms (like $\theta\zeta^2$ and $\theta^2\zeta^4$) will cancel in the second equation while other terms (like those without
$\zeta$) will cancel in third. Apparently, the resulting equations could be brought to a single equation, due to the symmetries of the system:

\begin{equation}
\begin{array}{l}
\theta \[ 2(p-1)(p-2)\zeta^3 - 2(p-1)(p-2q)\zeta^2 + 2(q-1)(q-2p)\zeta - 2(q-1)(q-2)   \] + (\zeta - 1) = 0.
\end{array} \label{eom2_3}
\end{equation}

One can see that the first and second, as well as third and fourth terms could be combined, but we need to add and subtract particular terms: $4(p-1)\zeta^2$ for the second term to take a form
$2(p-1)(p-2)\zeta^2$ and $4(q-1)\zeta$ for the third term to take a form $2(p-1)(p-2)\zeta$. Then

\begin{equation}
\begin{array}{l}
\theta \[ 2(p-1)(p-2)\zeta^2 (\zeta-1) + 2(q-1)(q-2)(\zeta-1) +4q(p-1)\zeta^2 - \right. \\ \left. - 4(p-1)\zeta^2 - 4p(q-1)\zeta + 4(q-1)\zeta   \] + (\zeta - 1) = 0; \\
\theta \[ 2(p-1)(p-2)\zeta^2 (\zeta-1) + 2(q-1)(q-2)(\zeta-1) +4(q-1)(p-1)\zeta^2  + 4(q-1)(p-1)\zeta   \] + \\ +  (\zeta - 1) = 0; \\
\theta \[ 2(p-1)(p-2)\zeta^2 (\zeta-1) + 2(q-1)(q-2)(\zeta-1) +4(q-1)(p-1)\zeta(\zeta-1)   \] + (\zeta - 1) = 0; \\
\theta \[ 2(p-1)(p-2)\zeta^2 + 2(q-1)(q-2) +4(q-1)(p-1)\zeta   \] + 1 = 0;
\end{array} \label{eom2_4}
\end{equation}

\noindent solving it with respect to $\theta$ gives us

\begin{equation}
\begin{array}{l}
\theta = - \dac{1}{2P_2}, ~~ P_2 = (p-1)(p-2)\zeta^2 + (q-1)(q-2) +2(q-1)(p-1)\zeta.
\end{array} \label{eom2_theta}
\end{equation}

Now we can substitute it into one of (\ref{eom2_2}) to obtain $\xi$:

\begin{equation}
\begin{array}{l}
\xi = - \dac{1}{4} \dac{P_1}{P_2^2}, ~~ P_1 = (p+1)p(p-1)(p-2)\zeta^4 + 4p(p-1)^2(q-1)\zeta^3 + \\ + 2(q-1)(p-1)(3pq-2p-2q)\zeta^2 + 4q(q-1)^2(p-1)\zeta + (q+1)q(q-1)(q-2).
\end{array} \label{eom2_xi}
\end{equation}

It is easy to confirm that $(3pq-2p-2q) > 0$ for $\{p, q\} \geqslant 2$, so that all the coefficients of $P_1$ are non-negative (though, it does not mean that the entire polynomial
$P_1$ is also non-negative---see detailed analysis below).
The discriminant of $P_1$ with respect to $\zeta$ is equal to

\begin{equation}
\begin{array}{l}
\mathcal{D}_{P_1} = -1024pq(p-1)^3(q-1)^3(p+q-1)(p+q-3)(p+q)^2 \times \\ \times (p^2q+pq^2-8pq+q^2+8p-q)(p^2q+pq^2-8pq+p^2+8q-p) < 0;
\end{array} \label{eom2_P1disc}
\end{equation}

\noindent one can verify that each of the multiplier is positive (under $\{p, q\} \geqslant 2$) and given minus sign in front, the discriminant is negative, which means there are two distinct real
roots---we will use this result later.

Let us turn our attention to $P_2$---its discriminant is equal to $\mathcal{D}_{P_2} = 4(q-1)(p-1)(p+q-3)$ and for $\{p, q\} \geqslant 2$ it is positive, resulting in two real roots:

\begin{equation}
\begin{array}{l}
\zeta_\pm = \dac{-(p-1)(q-1) \pm \sqrt{(q-1)(p-1)(p+q-3)}}{(p-1)(p-2)}.
\end{array} \label{eom2_denum_roots}
\end{equation}

\begin{figure}
\includegraphics[width=0.9\textwidth, angle=0]{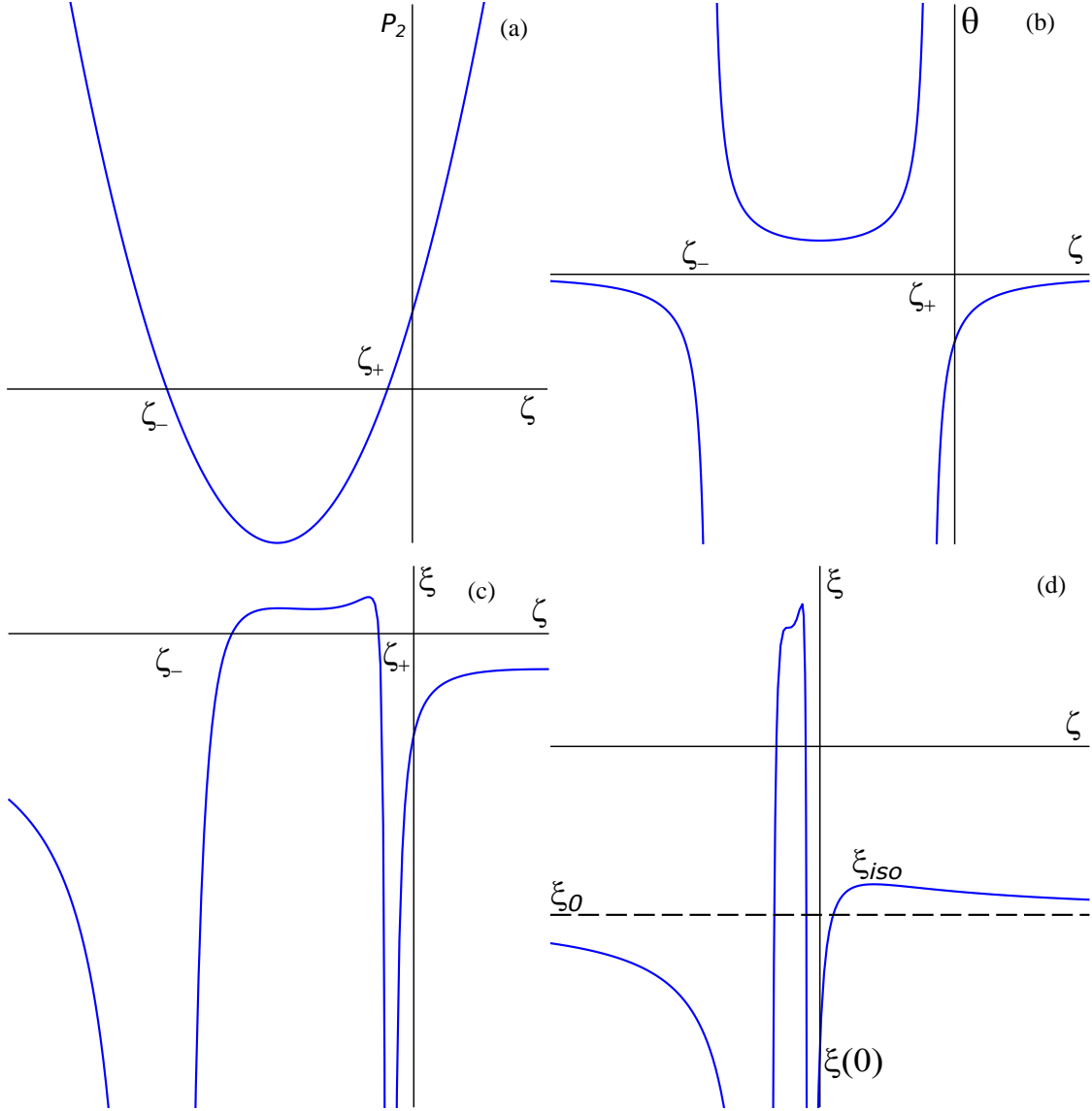}
\caption{Typical $P_2(\zeta)$ on panel (a) and $\theta(\zeta)$ on panel (b) (see Eq.(\ref{eom2_theta}) as well as $\xi(\zeta)$ on panel (c) (see Eq.(\ref{eom2_xi}) with location of roots $\zeta_\pm$
(see Eq.(\ref{eom2_denum_roots}). On (d) panel we presented large-scale $\xi(\zeta)$ behavior with $\zeta>0$ detailed part as well as $\zeta\to\pm\infty$ asymptote $\xi_0$ (\ref{eom2_xi0})
(see the text for more details).}\label{fig1}
\end{figure}

It is easy to see that $\zeta_- < 0$, as for $\zeta_+$, considering $(q-1)^2(p-1)^2 - (p-1)(q-1)(p+q-3) = \dots = (p-1)(p-2)(q-1)(q-2) \geqslant 0$, so that $\zeta_+ \leqslant 0$. Thus $P_2$ is a
parabola with both roots lying in $\zeta < 0$ (see Fig.~\ref{fig1}(a)) with $\zeta_\pm$ being its roots. Then with use of (\ref{eom2_theta}) we can derive that
$\theta$ looks like illustrated in Fig.~\ref{fig1}(b), and we can conclude that
$\theta\in(-\infty, 0)\cup(\theta_{min}, +\infty)$, with $\theta_{min} = \theta(\zeta_{min}) = \dac{p-2}{2(q-1)(p+q-3)}$.

To get $\xi$ we take square of $\theta$ and multiply by minus $P_1$ -- resulting graph is depicted in Fig.~\ref{fig1}(c). The range for $\xi$ is $\xi \leqslant \xi_{max}$, where $\xi_{max}$ is the
maximal possible
value found from investigation of structure of minima and maxima. Solving $d\xi/d\zeta = 0$ gives us three locations within $\zeta < 0$: $\zeta_1 = -(q-2)/(p-1)$, $\zeta_2 = -q/p$,
$\zeta_3 = -(q-1)/(p-2)$. There is one more extremum at $\zeta = 1$ and we shall return to it shortly.
One can easily check that $\zeta_1 - \zeta_3 = (p+q-3)/((p-1)(p-2)) > 0$ so that $\zeta_1 > \zeta_3$ always. As of the remaining differences,
$\zeta_1 - \zeta_2 = (2p-q)/(p(p-1))$, so that $\zeta_1 < \zeta_2$ if $q>2p$ and $\zeta_2 - \zeta_3 = (2q-p)/(p(p-2))$, so that $\zeta_2 < \zeta_3$ if $p>2q$.
Let us also note that all $\zeta_{1, 2, 3}$ are located within $\zeta_\pm$ -- this could be seen by considering all differences $\zeta_\pm - \zeta_{1, 2, 3}$, and their consideration shows that
$\zeta_- < \zeta_{1, 2, 3}$ and $\zeta_+ > \zeta_{1, 2, 3}$, which means that all $\zeta_{1, 2, 3}$ always located within $(\zeta_-, \zeta_+)$ range.

Let us go back to extremum analysis: to decide if we have maxima or minima
at $\zeta_{1, 2, 3}$, we need to find second derivatives there:

\begin{equation}
\begin{array}{l}
\xi''(\zeta_1) \equiv \left. \dac{d^2\xi}{d\zeta^2}\right|_{\zeta_1} = - \dac{2(q-1)(p-1)^2(2p-q)}{(q-2)^3(p+q-3)},~\xi''(\zeta_3) \equiv \left. \dac{d^2\xi}{d\zeta^2}\right|_{\zeta_3}
= \dac{2(p-1)(p-2)(p-2q)}{(q-1)^2(p+q-3)}, \\ \\
\xi''(\zeta_2) \equiv \left. \dac{d^2\xi}{d\zeta^2}\right|_{\zeta_2} = - \dac{2p^4(q-1)(p-1)(2p-q)(p-2q)(p+q)}{(p^2q+pq^2-2p^2+2pq-2q^2)^3}.
\end{array} \label{eom2_xi2der}
\end{equation}

It is easy to verify that all denominators $\xi''(\zeta_{1, 2, 3})$ are positive if $\{p, q\} > 2$, so that:

\begin{enumerate}

\item[I] for $p>2q$ we have $\zeta_2 < \zeta_3 < \zeta_1$ and we have maxima at $\zeta_2$, minima at $\zeta_3$ and maxima at $\zeta_1$;

\item[II] for $p<2q$ but $q<2p$ we have $\zeta_3 < \zeta_2 < \zeta_1$ and we have maxima at $\zeta_3$, minima at $\zeta_2$ and maxima at $\zeta_1$;

\item[III] finally for $q>2p$ we have $\zeta_3 < \zeta_1 < \zeta_2$ and we have maxima at $\zeta_3$, minima at $\zeta_1$ and maxima at $\zeta_2$.

\end{enumerate}

From this scheme (which is summarized in Fig.~\ref{fig2}(a)) one can see that the sequence ``maxima--minima--maxima'' preserves for all $\{p, q\} > 2$. Finally let us find this maximal possible value
for $\xi$; to do this we evaluate $\xi$ and all these points:

\begin{figure}
\includegraphics[width=0.9\textwidth, angle=0]{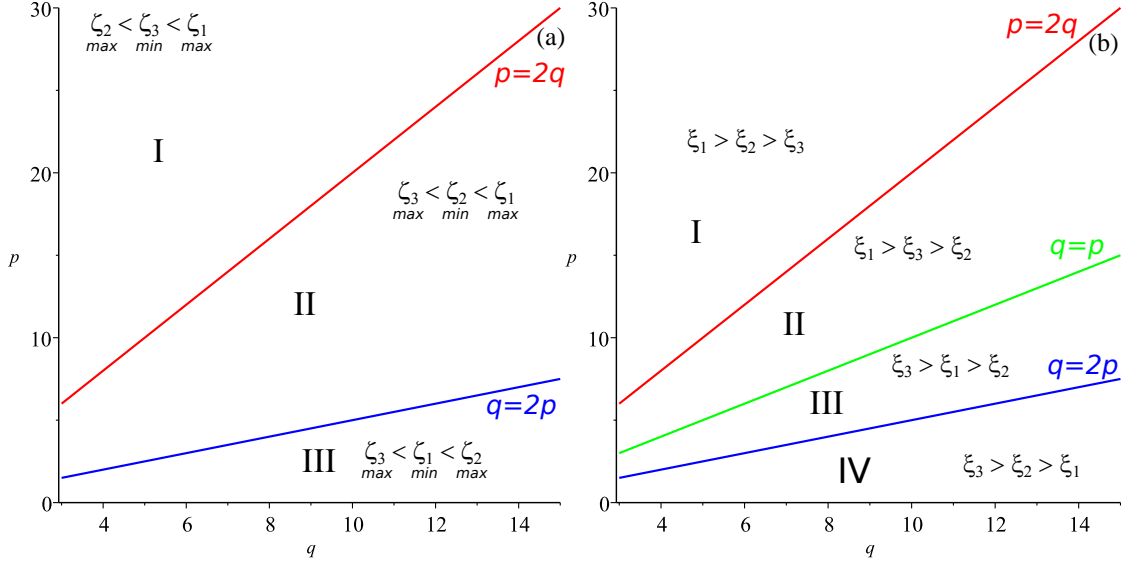}
\caption{Structure of maxima and minima among $\zeta_{\{1, 2, 3\}}$ on panel (a) and structure of $\xi_{max}$ among $\xi_{\{1, 2, 3\}}$ on panel (b) (see the text for more details).}\label{fig2}
\end{figure}

\begin{equation}
\begin{array}{l}
\xi_1 \equiv \xi(\zeta_1) = \dac{p^2q+pq^2-8pq+q^2+8p-q}{4(q-2)(p-1)(p+q-3)},~\xi_3 \equiv \xi(\zeta_3) = \dac{p^2q+pq^2-8pq+p^2+8q-p}{4(p-2)(q-1)(p+q-3)}, \\ \\
\xi_2 \equiv \xi(\zeta_2) = \dac{pq(p+q)}{4(p^2q+pq^2-2p^2-2q^2+2pq)},
\end{array} \label{eom2_ximinmax}
\end{equation}

\noindent and calculate their differences to compare them:

\begin{equation}
\begin{array}{l}
\xi_1 - \xi_2  = \dac{1}{2} \dac{(2p-q)^3(q-1)}{(p+q-3)(q-2)(p-1)(p^2q+pq^2-2p^2-2q^2+2pq)}, \\ \\ \xi_1 - \xi_3  =
\dac{1}{2} \dac{(p+q-3)(p-q)}{(p-1)(p-2)(q-1)(q-2)} \\ \\
\xi_2 - \xi_3  = \dac{1}{2} \dac{(p-2q)^3(p-1)}{(p+q-3)(p-2)(q-1)(p^2q+pq^2-2p^2-2q^2+2pq)}.
\end{array} \label{eom2_ximinmax}
\end{equation}

From differences we can see that we have new $p=q$ separatrix in addition to $p=2q$ and $q=2p$ from (\ref{eom2_xi2der}). Consideration of all possible $\{p, q\}$ cases gives us:

\begin{enumerate}

\item[I] for $p>2q$ we have $\xi_1 > \xi_2$, $\xi_1 > \xi_3$, $\xi_2 > \xi_3$, resulting in $\xi_1 > \xi_2 > \xi_3$;

\item[II] for $p<2q$ but $p>q$ we have $\xi_1 > \xi_2$, $\xi_1 > \xi_3$, $\xi_2 < \xi_3$, resulting in $\xi_1 > \xi_3 > \xi_2$;

\item[III] for $2p>q$ but $p<q$ we have $\xi_1 > \xi_2$, $\xi_1 < \xi_3$, $\xi_2 < \xi_3$, resulting in $\xi_3 > \xi_1 > \xi_2$;

\item[IV] and finally for $2p<q$ $\xi_1 < \xi_2$, $\xi_1 < \xi_3$, $\xi_2 < \xi_3$, resulting in $\xi_3 > \xi_2 > \xi_1$.

\end{enumerate}

So that the line $p=q$ separate $\xi_{max} = \xi_1$ for $p>q$ from $\xi_{max} = \xi_3$ for $p<q$; of course the transition is smooth:

\begin{equation}
\begin{array}{l}
\left. \xi_1\right|_{p=q} = \left. \xi_3\right|_{p=q} = \dac{q(2q^2-7q+7)}{4(q-1)(q-2)(2q-3)} > 0.
\end{array} \label{eom2_xismooth}
\end{equation}

This part of the scheme is illustrated in Fig.~\ref{fig2}(b). Let us also note that $\xi_{1, 3}$ decrease with growth of $\{p, q\}$ with $\lim_{\{p, q\}\to\infty} \xi_{1, 3} = 1/4$; general
function $\xi(\zeta)$, shown in  Fig.~\ref{fig1}(c), also has asymptotic behavior as $\zeta\to\pm\infty$:

\begin{equation}
\begin{array}{l}
\xi_0 = \lim\limits_{\zeta\to\pm\infty}\xi(\zeta) = - \dac{1}{4} \dac{p(p+1)}{(p-1)(p-2)};
\end{array} \label{eom2_xi0}
\end{equation}

\noindent this way it is defined for $p\geqslant q$; for $q\geqslant p$ it will be the same expression but with $p$ replaced by $q$. If replaced,
the resulting expression exactly coincide with $\xi(0)$---the value
of $\xi$ at $\zeta=0$ which separates the case when both subspaces have same behavior (both expand or both contract) from the case when subspaces have different behavior (one expands and one contracts):

\begin{equation}
\begin{array}{l}
\xi(0) = - \dac{1}{4} \dac{q(q+1)}{(q-1)(q-2)}.
\end{array} \label{eom2_xi00}
\end{equation}

\noindent One can see that $\xi_0$ and $\xi(0)$ swap places as $p\leftrightarrow q$. This situation as well as the behavior of $\xi(\zeta)$ for $\zeta>0$ is presented in Fig.~\ref{fig1}(d). One can also
note maximum at $\zeta=1$, mentioned earlier: $\zeta=1$ means that the expansion rate is the same for both subspaces meaning that it is isotropic solution; corresponding $\xi_{iso}$ is equal to

\begin{equation}
\begin{array}{l}
\xi_{iso} = - \dac{1}{4} \dac{(p+q)(p+q+1)}{(p+q-1)(p+q-2)};
\end{array} \label{eom2_xiiso}
\end{equation}

\noindent we shall discuss isotropic solution in details a bit below.

Finally let us find under which conditions the solutions exist: from Fig.~\ref{fig1}(b) we can see that for $\zeta\in(\zeta_-, \zeta_+)$ we have $\theta > 0$ while for
$\zeta\in(-\infty, \zeta_-)\cup(\zeta_+, +\infty)$ we have $\theta < 0$. Remembering that $\theta = \alpha h^2$ by definition, it is clear that for $\zeta\in(\zeta_-, \zeta_+)$ solutions exist only for
$\alpha > 0$ while for $\zeta\in(-\infty, \zeta_-)\cup(\zeta_+, +\infty)$ they exist only for $\alpha < 0$. From Figs.~\ref{fig1}(c)--(d) we can see that positive $\xi$ exist only within
$\zeta\in(\zeta_-, \zeta_+)$ range; for all other $\zeta$ we have $\xi < 0$. Remembering that $\xi = \alpha\Lambda$, we can summarize existence
conditions\footnote{Here by ``existence conditions'' we mean ``when the solutions {\it in principle} exist'', without specifying $\zeta$ (see a bit below).} as follows:

\begin{enumerate}

\item[] for $\alpha > 0$ solutions exist if $\Lambda \leqslant \xi_{max}/\alpha$ (including entire $\Lambda < 0$); resulting solutions have only $\zeta < 0$,

\item[] for $\alpha < 0$ solutions exist if $\Lambda \geqslant \xi_{iso}/\alpha$ (so that only $\Lambda > 0$); resulting solutions could have both signs for $\zeta$.

\end{enumerate}

Let us make one last note before summarizing the results: the scheme above formally does not work for $p=2q$ and $q=2p$, as for these conditions some of the second derivatives (\ref{eom2_xi2der})
nullify:

\begin{enumerate}

\item[] for $p=2q$ we have $\zeta_1 = -(q-2)/(2q-1)$, $\zeta_2 = \zeta_3 = -1/2$ and the second derivatives $\xi''(\zeta_2) = \xi''(\zeta_3) = 0$, $\xi''(\zeta_1) = -2q(2q-1)^2/(q-2)^3 < 0$, making
$\zeta_1$ only extremum, namely, maximum;

\item[] for $q=2p$ we have $\zeta_3 = -(2p-1)/(p-2)$, $\zeta_2 = \zeta_1 = -2$ and the second derivatives $\xi''(\zeta_2) = \xi''(\zeta_1) = 0$, $\xi''(\zeta_3) = -2p(p-2)/(2p-1)^2 < 0$, making
$\zeta_3$ only extremum, namely, maximum.

\end{enumerate}

So that in both cases instead of maximum-minimum-maximum structure we have degenerate structure with just one maximum instead; of course (this could be easily checked) in that case
$\xi_{max} = \xi(\zeta_{1, 2, 3})$, so that, formally, the structure could be treated as unchanged, just being degenerate at $p=2q$ and $q=2p$.

Described scheme could be summarized as follows: with given number of dimensions for both subspaces $p$ and $q$, as well as desired $\zeta = H/h$ (ratio of Hubble parameters acting in these
subspaces), using (\ref{eom2_theta}) we find $\theta = \alpha h^2$ and using (\ref{eom2_xi}) we find $\xi = \alpha\Lambda$. Now we have a degeneracy, as there are three variables ($\alpha$, $\Lambda$
and $h^2$) linked by two constraints, but this could be dealt with, say, by normalizing $\alpha = \pm 1$ (sign of $\alpha$ should coincide with that of $\theta$).
From this normalization we immediately have $|h|$ from $\theta$ and choose sign for $h$, then obtain $H$ with use of $\zeta$. From $\xi$ using the same normalized $\alpha$ as above we obtain $\Lambda$.
So as a result, we obtain $\alpha$ and $\Lambda$ for which out desired solution exist in $\{p, q\}$ dimensions and with Hubble parameters ratio $\zeta$.

Let us note that equations (\ref{eom2_theta}) and (\ref{eom2_xi}) always have solutions, so that
the proposed scheme allows one to {\it always} find a solution for a given $\zeta$ in $\{p, q\} > 2$ of interest---which means that, in any  $\{p, q\} > 2$ there {\it always} exist $\{\alpha, \Lambda\}$
such that there exist exponential solution with given $\zeta$. Let us also note that the direct procedure gives us single-valued functions---so that, for each set of $\zeta$, $\{p, q\} > 2$ and sign
for $\alpha$ there always exist single valued $\alpha$, $\Lambda$, $h$ and $H$ for the solution. But the inverse procedure is not necessary unique: starting from given $\alpha$ and $\Lambda$ we usually
end up with several exponential solutions which exist for these values of $\alpha$ and $\Lambda$.

Please mind that so far we were speaking about the existence of the exponential solution, not their stability, which we will discuss in the next subsection.

\section{Stability of the solutions}
\label{sec_stab}

It is important to know when the solutions exist, but it is also important to know the stability conditions as well. Indeed, solution could exist but if it is unstable, we cannot have it as an asymptotic
regime so it is ``useless'' in a way. Stability conditions for exponential solutions were studied in~\cite{my15, iv16, stab_add} and the criteria is formulated quite simple: $\sum H_i > 0$.
For our case it takes form

\begin{equation}
\begin{array}{l}
\sum_i H_i = pH + qh = h(p\zeta + q) > 0,
\end{array} \label{eom2_stab0}
\end{equation}

\noindent so that for $h>0$ we need $\zeta > \zeta_2$ (critical value for $\zeta$ from (\ref{eom2_stab0}) exactly coincide with $\zeta_2$ defined earlier) and for $h<0$ we need $\zeta < \zeta_2$.
So that the solution-finding scheme described earlier getting just a bit more complicated: the sign for $|h|$ is no longer our choice but is defined from input parameters $p$, $q$ and $\zeta$:
if $\zeta > -q/p$ then $h>0$, if $\zeta < -q/p$ then $h<0$; the rest of the scheme remains the same.

Let us consider stability conditions separately for $h>0$ and $h<0$.

\subsection{$h<0$ case}

For $h<0$ we need $\zeta < \zeta_2$, so that we have leftmost part in Fig.~\ref{fig1} and part of central (within $\zeta_\pm$) range, depending on $p$ and $q$. Within the central part we have
$\alpha > 0$ (since $\theta > 0$), while possible values for $\xi$ governed by $p$ and $q$:

\begin{enumerate}

\item for $q>2p$, $\zeta_2$ is located in the right (highest) maxima within $\zeta_\pm$ range and $\xi_{max}^- = \xi_2$;

\item for $q<2p$ but $p<2q$, $\zeta_2$ is located in minimum within $\zeta_\pm$ range and $\xi_{max}^- = \xi_3$;

\item for $p>2q$, $\zeta_2$ is located in the left (lower) maxima within $\zeta_\pm$ range and $\xi_{max}^- = \xi_2$;

\end{enumerate}

\noindent here by $\xi_{max}^-$ we denote maximal possible value for $\xi$ for stable solutions with $h<0$; below we comment on its relationship with $h>0$ counterpart.

So that the maximal possible value for $\xi$ depends on $p$ and $q$ but it is always positive, meaning that we can have $\xi$ of both signs (and so $\Lambda$ could be of both signs as well).

For $\zeta < \zeta_-$ we always have $\alpha < 0$ (since $\theta < 0$) and $\Lambda > 0$ (since $\xi < 0$), but $\xi$ is also limited from above by $\xi_0$, so that $\Lambda > \xi_0/\alpha$.

Summarizing, one can see that for $h<0$ we still can have solutions with both signs of $\alpha$: for $\alpha > 0$ with both signs of $\Lambda$ with $\Lambda < \xi_{max}^-/\alpha$
(including entire $\Lambda < 0$), while for $\alpha < 0$ only with $\Lambda > \xi_0/\alpha > 0$.
Solutions with $\Lambda < 0$ exist (and stable) only for $\alpha > 0$ and only for a narrow range of $\zeta$, but they cover the entire $\Lambda < 0$ range. Overall, stability and existence
criteria for $h<0$ resemble just existence criteria but different bounds---we shall comment on it below.

\subsection{$h>0$ case}

Analysis for $h>0$ is quite similar to the previous case: now we need $\zeta > \zeta_2$ and we have rightmost part in Fig.~\ref{fig1} and part of central (within $\zeta_\pm$) range, depending on
$p$ and $q$. Within the central region we have $\alpha > 0$ (since $\theta > 0$) and possible values for $\xi$ governed by $p$ and $q$:

\begin{enumerate}

\item for $q>2p$, $\zeta_2$ is located in the right (highest) maxima within $\zeta_\pm$ range and $\xi_{max}^+ = \xi_2$;

\item for $q<2p$ but $p<2q$, $\zeta_2$ is located in minimum within $\zeta_\pm$ range but $\xi_{max}^+ = \xi_1$;

\item for $p>2q$, $\zeta_2$ is located in the left (lower) maxima within $\zeta_\pm$ range and $\xi_{max}^+ = \xi_1$;

\end{enumerate}

\noindent here by $\xi_{max}^+$ we denote maximal possible value for $\xi$ for stable solutions with $h>0$.
Similarly to the previous case, maximal possible value for $\xi$ depends on $p$ and $q$ and it is again always positive, so that we can have $\xi$ (and so $\Lambda$) of both signs. Also similar to
the previous case, $\Lambda < 0$ solution exist and stable only for $\alpha > 0$ and only for a narrow range of $\zeta$, and again they cover the entire $\Lambda < 0$ range.

For $\zeta > \zeta_+$ we always have $\alpha < 0$ (since $\theta < 0$) and $\Lambda > 0$ (since $\xi < 0$), but $\xi$ is now limited from above by $\xi_{iso}$, so that $\Lambda > \xi_{iso}/\alpha$.

\subsection{Stability summary}

Summarizing, for both $h<0$ and $h>0$ we can have solutions with both signs of $\alpha$: for $\alpha > 0$ with both signs of $\Lambda$ while for $\alpha < 0$ only with $\Lambda > 0$. But
according to the analysis above, the limits are different: for $\alpha > 0$, $\Lambda > 0$ we have $\xi \equiv \alpha\Lambda \leqslant \xi_{max}^\pm$ where ``+'' sign corresponds to $h>0$ while
``--'' sing to $h<0$; note that $\xi_{max}^+ \geqslant \xi_{max}^-$ (see above) and we have $\xi_{max}^+ = \xi_{max}^-$ only if  $q>2p$.

For $\alpha < 0$, $\Lambda > 0$ domain limits are also different: for $h<0$ we have the leftmost part of Fig.~\ref{fig1}(d) and so $\xi < \xi_0$ while for $h>0$ we have the rightmost part of
Fig.~\ref{fig1}(d) and so $\xi < \xi_{iso}$; additionally $\xi_{iso} > \xi_0$.

\begin{figure}
\includegraphics[width=0.9\textwidth, angle=0]{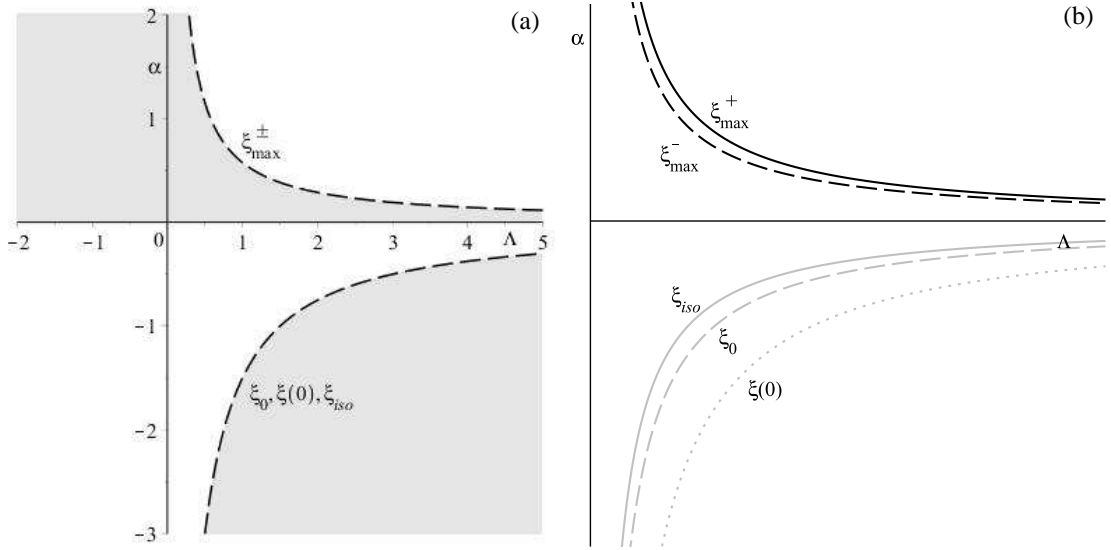}
\caption{Existence and stability areas on $(\alpha, \Lambda)$ plane for the general $\{p, q\}\geqslant 2$ case: general structure on (a) panel and details with locations of
different $\xi_{max}^\pm$ as well as $\xi_0$, $\xi(0)$ and $\xi_{iso}$ on (b) panel (see the text for more details).}\label{fig6}
\end{figure}

This structure as well as relationship between different limits are illustrated in Fig.~\ref{fig6}. There on (a) panel we presented general structure---it is the same for the existence (in this case we
use just $\xi_{max}$ for $\alpha > 0$, $\Lambda > 0$ and $\xi_{iso}$ for $\alpha < 0$, $\Lambda > 0$) and for stability. Details on how the bounds are modified are presented on (b) panel---there we presented
particular case $p=7$, $q=4$ (just as a way of example) and one can see there how $\xi_{max}^\pm$ differ from each other, as well as how $\xi_{iso}$ (which we use for $h>0$) differs from $\xi_0$
(which we use for $h<0$).

In addition to $\xi_0$ and $\xi_{iso}$ we presented $\xi(0)$ in Fig.~\ref{fig6}(b). We did it for the following reason: depending on the additional requirements, which could rise from the actual problem
to solve, the bounds defined above could tighten or loosen. And $\xi(0)$ illustrates one of such additional conditions---there could be a situation when we absolutely require $\zeta < 0$ (for instance,
we demand that one of the subspaces expand and another contract like in realistic compactification scheme, presented below in Section~\ref{sec_3D}),
in that case we consider only $\zeta < 0$ and so the bounds on $\xi$ modify
accordingly. Of course this is just an example (though a useful one), but, as we mentioned, depending on a specific problem at hand there could be other additional bounds and we need to describe how to handle
them.

\section{Special cases}
\label{sec_spec}

Through the analysis one could notice the we almost always have terms like $(p-1)$, $(p-2)$, $(q-1)$ and $(q-2)$ in either numerators or denominators of the expressions. That is why we stated that the results obtained are valid for $\{p, q\}>2$. Then it is natural to consider these cases with $\{p, q\}=\{1, 2\}$ separately.

\subsection{$\{p, q\}=1$ case}

This is a trivial case of $(1+1)$-dimensional gravity and it is not defined within Gauss-Bonnet gravity since the corresponding Euler invariant is identically zero.

\subsection{$q=1$ case}

For this case to be properly defined, we need $p\geqslant 3$. The equations of motion (\ref{eom2}) take a form

\begin{equation}
\begin{array}{l}
p(p-1)H^2 + 2(p-1)Hh + 2h^2 + \alpha\( p(p-1)(p-2)(p-3)H^4 + 4(p-1)^2(p-2)H^3h + \right. \\ \left. + 4(p-1)(p-2)H^2h^2  \)=\Lambda, \\
p(p+1)H^2 + \alpha p(p-1)(p-2)(p+1)H^4 = \Lambda, \\
p(p-1)H^2 + 2pHh + \alpha \( p(p-1)(p-2)(p-3)H^4 + 4p(p-1)(p-2)H^3h   \) = \Lambda.
\end{array} \label{eom_Dp1}
\end{equation}

One immediately can see that the middle equation does not have $h$ so it is valid for any $h$; apparently, the same is true for the entire system---indeed, the system has the only nontrivial  solution

\begin{equation}
\begin{array}{l}
\Lambda = \dac{1}{2} p(p+1)H^2, ~~ \alpha = - \dac{1}{2(p-1)(p-2)H^2}.
\end{array} \label{eom_Dp1_sol}
\end{equation}

One can easily verify that the solution is valid by direct substitution, and one can see that the solution does not depend on $h$, so it is valid for any $h$. This situation is surely unphysical; this fact
 was noted several times---first in~\cite{CPT1}, then while studying stability issues in~\cite{my15} and finally in~\cite{my16b} it becomes apparent that this solution is not part of the evolution.
It is interesting to note that if we use (\ref{eom2_2}) instead (the system rewritten in $\theta$, $\zeta$ and $\xi$), we would obtain formally valid solution

\begin{equation}
\begin{array}{l}
\theta = - \dac{1}{2(p-1)(p-2)\zeta^2}, ~~\xi = - \dac{1}{4} \dac{p(p+1)}{(p-1)(p-2)}.
\end{array} \label{eom_Dp1_for.sol}
\end{equation}

One can also see that under additional condition $h(p\zeta+1)>0$ obtained solution is formally stable but, as we already mentioned, it is unphysical and cannot be reached from the general
Einstein-Gauss-Bonnet cosmology (see~\cite{my16b} for discussions).

\subsection{$\{p, q\}=2$ case}
\label{q2p2subsec}

In this case the system (\ref{eom2_2}) takes form

\begin{equation}
\begin{array}{l}
2\theta\zeta^2(1+12\theta) + 8\theta\zeta + 2\theta = \xi, \\
2\theta\zeta^2(1+4\theta) + 4\theta\zeta(1+4\theta) + 6\theta = \xi, \\
16\theta^2\zeta^3 + 2\theta\zeta^2(3+4\theta) + 4\theta\zeta + 2\theta  \xi,
\end{array} \label{eom_2p2}
\end{equation}

\noindent and its solution is

\begin{equation}
\begin{array}{l}
\theta = -\dac{1}{4\zeta},~~\xi=-\dac{\zeta^2+\zeta+1}{2\zeta}.
\end{array} \label{eom_2p2_sol}
\end{equation}

One can see that the sign of $\alpha$ depends on the sign $\zeta$:

\begin{enumerate}

\item[] solutions with $\zeta < 0$ exist for $\alpha > 0$ and $\Lambda > 0$,

\item[] solutions with $\zeta > 0$ exist for $\alpha < 0$ and $\Lambda > 0$;

\end{enumerate}

\noindent the sign for $\Lambda$ is the same in both cases. Stability conditions for this case read: $\sum H_i = (2H + 2h) = h(2\zeta + 2) > 0$, so that for $h > 0$ solutions to be stable we need
$\zeta > -1$ while for $h<0$ solutions to be stable we need $\zeta < -1$.

\subsection{$q=2$ case}
\label{q2subsec}

For this case to be properly defined, we need $p\geqslant 2$, but since we already considered $p=2$ case, we demand $p>2$ here. The equations of motion (\ref{eom2_2}) take a form

\begin{equation}
\begin{array}{l}
p(p-1)(p-2)(p-3)\theta^2\zeta^4 + 8p(p-1)(p-2)\theta^2\zeta^3 + p(p-1)\theta(1+12\theta)\zeta^2 + 4p\theta\zeta+2\theta=\xi, \\
p(p-1)(p-2)(p-3)\theta^2\zeta^4 + 8(p-1)^2(p-2)\theta^2\zeta^3 + \theta(p-1)(20p\theta+p-32\theta)\zeta^2 + \\ + 4\theta(1+4\theta)(p-1)\zeta + 6\theta = \xi, \\
p(p-1)(p-2)(p+1)\theta^2\zeta^4 + 4p^2(p-1)\theta^2\zeta^3 + p\theta(4p\theta+p-4\theta+1)\zeta^2 + 2p\theta\zeta + 2\theta = \xi.
\end{array} \label{eom_Dp2}
\end{equation}

Solution of this system is

\begin{equation}
\begin{array}{l}
\theta = - \dac{1}{2\zeta(p-1)(\zeta(p-2)+2)}, ~~ \xi = - \dac{P_1}{4\zeta(p-1)(\zeta(p-2)+2)^2}, \\
P_1 = p(p+1)(p-2)\zeta^3 + 4p(p-1)\zeta^2 + 8(p-1)\zeta + 8.
\end{array} \label{eom_Dp2_sol}
\end{equation}

\begin{figure}
\includegraphics[width=0.95\textwidth, angle=0]{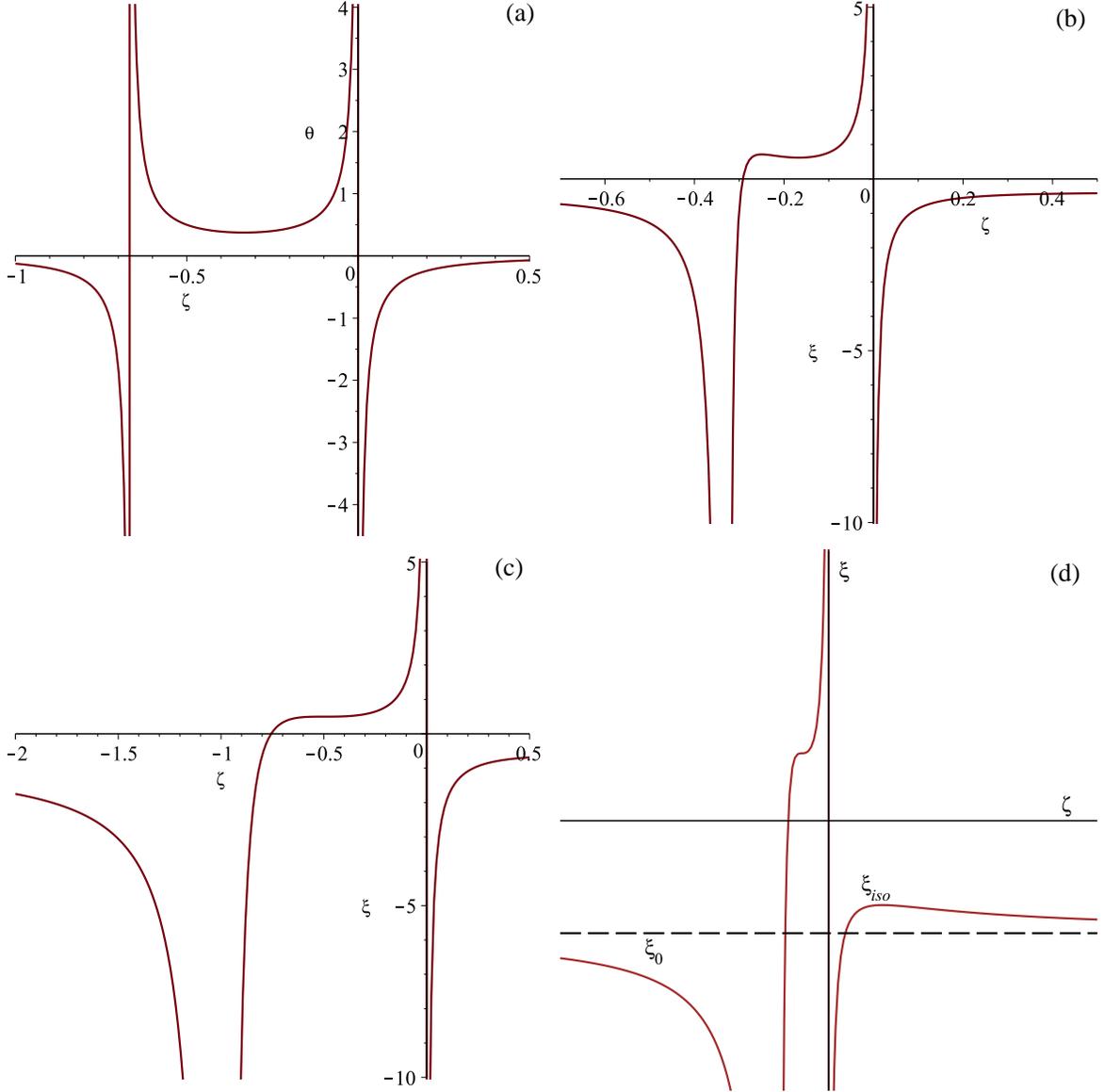}
\caption{Structure of the solutions for $q=2$ case: $\theta(\zeta)$ on (a) panel, $\xi(\zeta)$ for $p\ne 4$ on (b), for $p=4$ on (c) and ``large-scale'' structure on (d) panels
(see the text for more details).}\label{fig3}
\end{figure}

Comparing expression for $\theta$ with the general case (\ref{eom2_theta}) we can see that the denominators both are quadratic functions, but now we effectively have $\zeta_+ = 0$ and
$\zeta_- = -2/(p-2)$. Let us also note that unlike the general case, now we have $\lim_{\zeta\to\pm\infty} \theta = 0$ (see Fig.~\ref{fig3}(a)).

Comparison of the expressions for $\xi$ reveals substantial difference---in the general case (\ref{eom2_xi}) we have quartic function in the numerator while in $q=2$ case
it is cubic. Solving $d\xi/d\zeta = 0$ gives us two extrema for $\zeta < 0$: $\zeta_1 = -2/p$ and $\zeta_2 = -1/(p-2)$, and one more at $\zeta=1$, similar to the general case.
Evaluating second derivatives at $\zeta_{1, 2}$ yields

\begin{equation}
\begin{array}{l}
\xi''(\zeta_1) = - \dac{p^4(p+2)(p-4)}{128(p-1)}, ~~\xi''(\zeta_2) = 2(p-2)(p-4).
\end{array} \label{eom_Dp2_ddxi}
\end{equation}

One immediately can see multiplier $(p-4)$, meaning that $p=4$ separate two cases: for $p>4$ we have maximum at $\zeta_1$ and minimum at $\zeta_2$ while for $p<4$ we have maximum at $\zeta_2$ and
minimum at $\zeta_1$. But we also can notice that for $p>4$ we have $\zeta_2 > \zeta_1$ while for $p<4$ it is $\zeta_2 < \zeta_1$, so that the structure of extrema is the same for both $p>4$ and
$p<4$; at $p=4$ we have $\zeta_2=\zeta_1$---they coincide and form a saddle point. The situation is presented in Figs.~\ref{fig3}(b, c): on (b) panel we presented $p \ne 4$ case while on (c) panel
$p=4$. One can also see that unlike the general case, now $\xi(\zeta)$ is unbound both from below and from above. Finally, in Fig.~\ref{fig3}(d) we presented ``large-scale structure'' and, similar
to the general case, we again have $\xi_0$ and $\xi_{iso}$; their expressions coincide with those from the general case---(\ref{eom2_xi0}) and (\ref{eom2_xiiso}), respectively. Let us also note that,
unlike the general case, now $\xi(0)$ is undefined as we have $\zeta_+=0$ effectively.

\begin{figure}
\includegraphics[width=0.95\textwidth, angle=0]{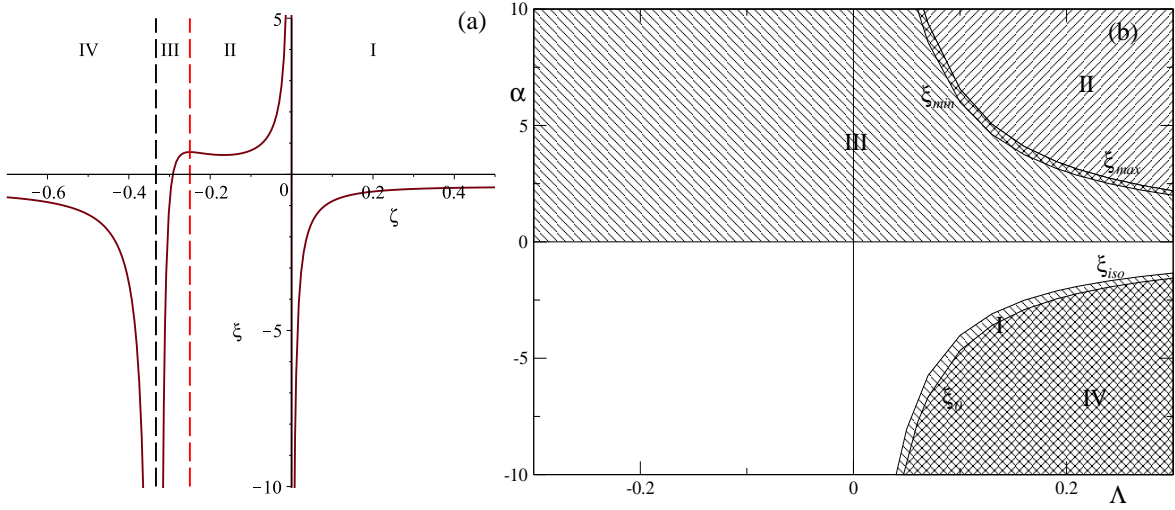}
\caption{Stability analysis for $q=2$ case: structure of the stable solutions on (a) panel and location of existing and stable regions within $\{\alpha, \Lambda\}$ plane on (b)
(see the text for more details).}\label{fig31}
\end{figure}

Let us now address the stability of the solutions. The criteria reads
$\sum H_i = (pH + 2h) = h(\zeta p + 2) > 0$, so that either $h > 0$ and $\zeta > -2/p$ or $h < 0$ and $\zeta < -2/p$. Let us note that the critical value for $\zeta$ coincide with $\zeta_1$ found
earlier which corresponds to the local maximum of $\xi(\zeta)$; its location is depicted with red dashed line in Fig.~\ref{fig31}(a); with black dashed line on the same graph we depicted $\zeta_-$.
So that the stability conditions could be summarized as follows:

\begin{enumerate}

\item[] solutions with $h > 0$ exist and stable iff $\zeta > -2/p$ which corresponds to regions I and II in Fig.~\ref{fig31}(a): in region I we have $\alpha < 0$, $\xi < \xi_{iso} < 0$ and so
$\Lambda > \xi_{iso}/\alpha > 0$; in region II we have $\alpha > 0$, $\xi > \xi_{min}$ so that $\Lambda > \xi_{min}/\alpha > 0$;

\item[] solutions with $h < 0$ exist and stable iff $\zeta < -2/p$ which corresponds to regions III and IV in Fig.~\ref{fig31}(a): in region III we have $\alpha > 0$, $\xi < \xi_{max}$
(including entire $\xi < 0$) so that $\Lambda < \xi_{max}/\alpha$ (including entire $\Lambda < 0$); in region IV we have $\alpha < 0$, $\xi < 0$ and so $\Lambda > 0$.

\end{enumerate}

Mentioned above $\xi_{\{min, max\}}$ could be exactly found:

\begin{equation}
\begin{array}{l}
\xi_{min, max} =  \min, \max (\xi_1, \xi_2),    ~~
\xi_1 = \dac{p(p+2)}{16(p-1)}, ~~\xi_2 = \dac{3p^2 - 13p + 16}{4(p-1)(p-2)}.
\end{array} \label{eom_Dp2_ximinmax}
\end{equation}

The entire situation is illustrated in Fig.~\ref{fig31}(b), where we transferred regions from Fig.~\ref{fig31}(a) to corresponding areas in $\{\alpha, \Lambda\}$ plane; ordering of the regions is the same
on both panels. Regions I and IV are located in the fourth quadrant and are bounded by $\xi_{iso}$ and $\xi_0$ respectively; since $\xi_{iso} > \xi_0$, double-shaded region is region IV. Region II
is located in the first quadrant and is bounded from below by $\xi_{min}$ while Region III is located in the first and second quadrants; it is bounded by $\xi_{max}$ from above in first quadrant.
Again, since $\xi_{max} > \xi_{min}$
(except $q=4$ when they coincide), there exist double-shaded area where solutions from both regions exist and are stable.

So that one can see that solutions with $h>0$ (Regions I and II in Fig.~\ref{fig31}(b)) exist and stable only under $\Lambda > 0$ while those with $h<0$ (Regions III and IV in Fig.~\ref{fig31}(b))
could be stable with $\Lambda < 0$, but this is happening for a very narrow range of $\zeta$.

Concluding, the $q=2$ case mostly follows the general scheme, with two main differences:

\begin{enumerate}

\item[i] $\xi(\zeta)$ curve is unbounded from above within $\zeta_\pm$ which results in different bounds on $\xi$ (compare Fig.~\ref{fig6}(a) with Fig.~\ref{fig31}(b));

\item[ii] the expressions for $\zeta_\pm$ are different, in particular, $\zeta_+=0$, which results in undefined $\xi(0)$.

\end{enumerate}

One can consider $q=2$ scheme as ``oversimplified'' general scheme for the following reason: as we already set $q=2$, we cannot have cases like $p<2q$, and only $p>q$ subcases remains. For this reason,
separation point $\zeta_2$ already fixed and we do not have this variation either.

\subsection{Isotropic solution}
\label{isosubsec}

Last special case we want to consider is the isotropic solution. It is a situation when the entire space is isotropic and expands exponentially. Obviously, this corresponds to the case $\zeta=1$, and
we shall use both direct approach and substitution of $\zeta=1$ into derived equations.

For direct approach we use general set of equations (\ref{eom2}) and substitute $h=H$ as well as $(p+q)=D$; then the system collapse to a single equation

\begin{equation}
\begin{array}{l}
\alpha D(D-1)(D-2)(D-3)H^4 + D(D-1)H^2 = \Lambda.
\end{array} \label{eom2_iso}
\end{equation}

This is biquadratic equation with respect to $H$ and it could be exactly solved:

\begin{equation}
\begin{array}{l}
H^2_\pm = \dac{-D(D-1)\pm\sqrt{4(D-2)(D-3)\xi + D(D-1)}}{2\alpha D(D-1)(D-2)(D-3)} = \\ \\ = \dac{-D(D-1)\pm\sqrt{D(D-1)(D-2)(D-3)(\xi - \xi_{iso})}}{2\alpha D(D-1)(D-2)(D-3)},
\end{array} \label{eom2_iso_sol1}
\end{equation}

\noindent where $\xi_{iso}$ is exactly the same as in Eq.~(\ref{eom2_xiiso}).

Now let us see if we could obtain similar result from our approach. First of all, let us note that according to (\ref{eom2_iso}) there could be up to two distinct solutions, but our intermediate equation
(\ref{eom2_3}) is already linear in $\theta$. This is happening because to obtain it we use difference equations obtained from (\ref{eom2_2}). This way we get rid of $\xi$, but also from one of the powers
of $\theta$, effectively loosing one of the isotropic solutions. But if we substitute $\zeta=1$ into (\ref{eom2_2}), all three equations become the same:

\begin{equation}
\begin{array}{l}
(p+q)(p+q-1)(p+q-2)(p+q-3)\theta^2 + (p+q)(p+q-1)\theta = \xi.
\end{array} \label{eom2_iso2}
\end{equation}

One can see that in this form the equation is still quadratic and, under $(p+q)=D$ and definitions for $\theta$ and $\xi$ exactly coincide with (\ref{eom2_iso}). So that on the level of equations, direct
approach and our representation gives the same results; though, the general course leads to loose of one of the isotropic roots.

\section{Realistic compactification}
\label{sec_3D}

Finally let us address one more important case---namely, the case with three-dimensional subspace. This case could describe the final stage of generic compactification---natural scenario with
expanding three and contracting extra dimensions. Formally this case fall within general scheme, but due to its utmost relevance to the physical cosmology we decided to consider it separately.

So we set $q=3$ and demand $h>0$ (three-dimensional subspace---``our Universe''---is expanding, as it is known) and $\zeta < 0$ (so that $H<0$ and so extra dimensions are contracting---that is why we
cannot detect them by any means); the equations of motion take form

\begin{equation}
\begin{array}{l}
p(p-1)(p-2)(p-3)\theta^2\zeta^4 + 12p(p-1)(p-2)\theta^2\zeta^3 + p(p-1)\theta(1+36\theta)\zeta^2 + 6p\theta(4\theta+1)\zeta + 6\theta=\xi, \\
p(p-1)(p-2)(p-3)\theta^2\zeta^4 + 12(p-1)^2(p-2)\theta^2\zeta^3 + \theta(p-1)(48p\theta + p - 72\theta)\zeta^2 + \\ + 6\theta(1+12\theta)(p-1)\zeta + 12\theta(2\theta+1) = \xi, \\
p(p-1)(p-2)(p+1)\theta^2\zeta^4 + 8p^2(p-1)\theta^2\zeta^3 + p\theta(20p\theta+p-12\theta+1)\zeta^2 + 4p\theta(1+4\theta)\zeta + 6\theta = \xi.
\end{array} \label{eom_Dp3}
\end{equation}

One can see that for $p=3$ the system simplifies---we will comment on it later; for now we assume $p>3$ so that all the terms give nontrivial contribution. The expressions for $\theta$ and $\xi$
could be obtained from (\ref{eom2_theta}) and (\ref{eom2_xi}) by substituting $q=3$, then $d\xi/d\zeta = 0$ gives us $\zeta_1 = -1/(p-1)$, $\zeta_2 = -3/p$, $\zeta_3 = -2/(p-2)$. If we try to apply
the scheme for minima and maxima structure within $\zeta_\pm$, we would notice the following: $q=2p$ is never reached (since it corresponds to $p=3/2$ which violates $p>3$ condition while
$p=2q$ gives us $p=6$. So that for $\zeta$'s structure (see Fig.~\ref{fig2}(a)) we are limited to I and II regions while for maximal $\xi$ (see Fig.~\ref{fig2}(b)) we are also limited to I and II
regions (please note that the regions structure is a bit different in Figs.~\ref{fig2}(a) and (b)).

Now let us add stability condition: $\sum H_i = (pH+3h) = h(p\zeta+3) > 0$. We demand $h>0$ (so that ``our Universe''---three-dimensional submanifold---expands) so that we need $(p\zeta +3) > 0$, which
results in $\zeta > -3/p$, which is exactly $\zeta_2$. But we also require $\zeta < 0$, so that extra dimensions to have $H<0$ -- we want them to contract. Then, combining the two, we obtain
$0 > \zeta > \zeta_2$. This region could be splitted into three parts, divided by $\zeta_+$ and some $\zeta_0$, larger root of $P_1 = 0$. Then we report existence and stability within each of the regions:

\begin{enumerate}

\item[] for $\zeta\in(\zeta_+, 0)$ we have $\theta < 0$ which results in $\alpha < 0$, and $\xi < \xi(0) = -3/2$, so that $\xi$ is negative and so $\Lambda > 0$. So that the solutions with
$\zeta\in(\zeta_+, 0)$ exist for $\alpha < 0$, $\Lambda > 0$, $\alpha\Lambda < -3/2$;

\item[] for $\zeta\in(\zeta_0, \zeta_+)$ we have $\theta > 0$ which results in $\alpha > 0$, and $\xi < 0$, which results in $\Lambda < 0$. Let us note that the entire $\xi < 0$ is covered within
$\zeta\in(\zeta_0, \zeta_+)$, so that for each $(\alpha, \Lambda)$ with $\alpha > 0$ and $\Lambda < 0$ there exist stable anisotropic exponential solution with $\zeta\in(\zeta_0, \zeta_+)$;

\item[] for $\zeta\in(\zeta_2, \zeta_0)$ we have $\theta > 0$ which results in $\alpha > 0$, and $\xi_{max} > \xi > 0$, which results in $\Lambda > 0$. So that for
each $(\alpha, \Lambda)$ with $\alpha > 0$ and $\Lambda > 0$ ($\alpha\Lambda < \xi_{max}$) there exist stable anisotropic exponential solution with $\zeta\in(\zeta_2, \zeta_0)$.

\end{enumerate}

There remain a few things to do to complete the description. In the section dedicated to the general structure we described swapping between $\zeta$'s
depending on $p$ and $q$. Now we already have $q=3$ and have restriction on $p>3$. This leave us with only I and II regions in Figs.~\ref{fig2}(a, b), and they interchanged at $p=2q = 6$. So that
at $p<6$ we have $\zeta_2$ located at the local minima and at $p>6$ we have $\zeta_2$ located at the local maxima; at $p=6$ they coincide, forming a saddle point -- exactly as in the general case.
Nevertheless, in both cases global maximum is included within $\zeta\in(\zeta_2, 0)$; according to the general scheme it is $\xi(\zeta_1)$ and it is equal to

\begin{equation}
\begin{array}{l}
\xi_{max} = \xi\(\zeta_1 = - \dac{1}{p-1}\) = \dac{3p^2-7p+6}{4p(p-1)} > 0;
\end{array} \label{eom_Dp3_ximax}
\end{equation}

\noindent this result coincide with one obtained earlier in~\cite{my17a}.

Described above structure of the stable solutions is illustrated in Fig.~\ref{fig4}(a). By vertical black dashed lines we presented positions of $\zeta_2$ (which is left bounding value for stable solution)
for two cases: $p>6$ and $p<6$; for $p=6$ they coincide. Vertical red dashed line is $\zeta_+$ which separate region with $\alpha < 0$ (for $\zeta > \zeta_+$) from $\alpha > 0$ (for $\zeta < \zeta_+$).
By $\zeta_0$ we denoted larger root of $P_1 = 0$:

\begin{figure}
\includegraphics[width=0.9\textwidth, angle=0]{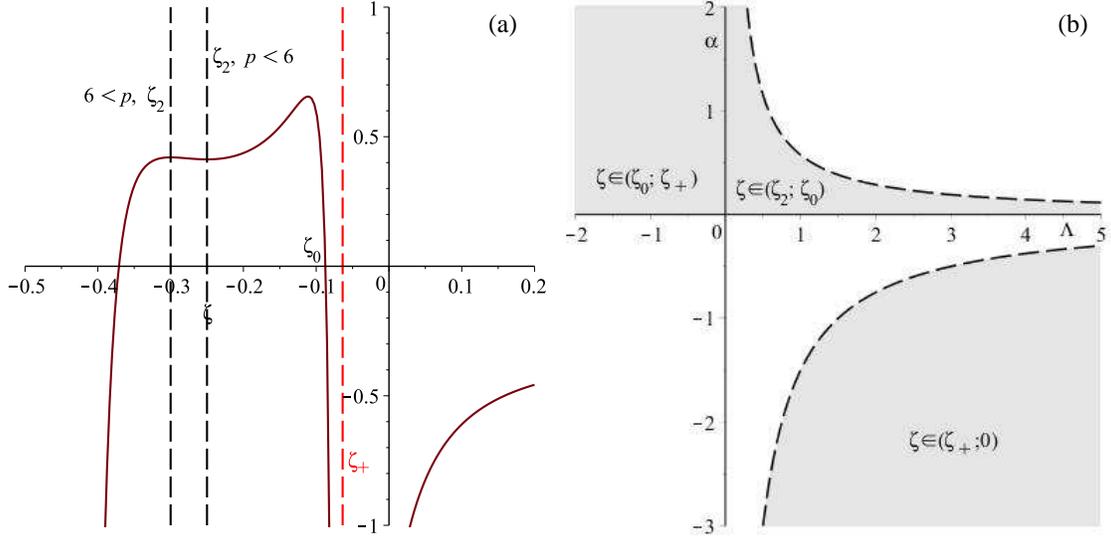}
\caption{Structure of the solutions for $q=3$ case with realistic compactification on (a) panel and distribution of all possible $\zeta$'s over $\{\alpha, \Lambda\}$ plane for $q=3$ stable solutions
(see the text for more details).}\label{fig4}
\end{figure}

\begin{equation}
\begin{array}{l}
P_1 = p(p-1)(p-2)(p+1)\zeta^4 + 8p(p-1)^2\zeta^3 + 4(7p-6)(p-1)\zeta^2 + 48(p-1)\zeta + 24.
\end{array} \label{eom_Dp3_P1}
\end{equation}

\noindent The discriminant of this equation with respect to $\zeta$

\begin{equation}
\begin{array}{l}
\mathcal{D} = -98304p^2(p-1)^3(p+2)(p+3)^2(3p^2-7p+6)(p^2-4p+6) < 0,
\end{array} \label{eom_Dp3_P1D}
\end{equation}

\noindent so that it always has 2 distinct real roots.

Now let us comment on $p=3$ -- as one can see from (\ref{eom_Dp3}), equations simplify in this case, but keeping in mind that only two of them are independent and one of these two independent is the last one,
which is still $\zeta^4$ for $p=3$. Another independent equation could be obtained as a difference between first and second and, even for $p>3$, first term is nullified anyway, so that the resulting form
of the equation is the same ``simplified'' both for $p=3$ and $p>3$ -- so that our results are the same in $p=3$ case as well.

This finalize our study of $q=3$ cases. We saw that the structure of the solutions follow the general scheme, but since one of the dimensions is fixed ($q=3$) and from physical considerations we
also have limitations on $h>0$ and $\zeta<0$, the resulting solution is much more simple---in this regard the reasoning is similar to the $q=2$ case (like, we cannot have $q>2p$ subcase and have only
$p\geqslant q$ etc.). We have proven that the stable solutions with compactification exist for $\alpha > 0$,
$\Lambda < \xi_{max}/\alpha$ (including entire $\Lambda < 0$) as well as $\alpha < 0$, $\Lambda > 0$, $\alpha\Lambda < -3/2$ -- these areas are combined in Fig.~\ref{fig4}(b) together with
corresponding $\zeta$'s.

\section{Summary and discussions}

In this paper we proposed a scheme which allows to find anisotropic exponential solution in EGB gravity with metric being a product of two subspaces with dimensionalities $p$ and $q$. We choose
$\zeta \equiv H/h$ the desired ratio of Hubble parameters and with all three quantities ($p$, $q$ and $\zeta$) calculate $\theta = \alpha h^2$ with use of (\ref{eom2_theta}) and $\xi = \alpha\Lambda$
with use of (\ref{eom2_xi}). Now we formally have a degeneracy as we have three parameters ($\alpha$, $\Lambda$ and $h^2$) linked by two equations, but this situation could be dealt with, say, by
choosing normalization for $\alpha = \pm 1$, with the sign for $\alpha$ determined by the resulting sign for $\theta$. With the degeneracy being dealt with, we finally find $h^2$ and $\Lambda$, choose
sign for $h$ and immediately find $H$ from it and given $\zeta$. Overall, from given $p$, $q$ and $\zeta$ we obtain $H$ and $h$ as well as
$\alpha$ (normalized) and $\Lambda$ for which this solution exist. The equations (\ref{eom2_theta}) and (\ref{eom2_xi}) always have solutions (except zeros of denominator) so that we can claim that
except for these singular points for $\zeta$ for any $\{p, q\} > 2$ there always exist $\{\alpha, \Lambda\}$ such that the solution in question exists.

\begin{figure}
\includegraphics[width=0.9\textwidth, angle=0]{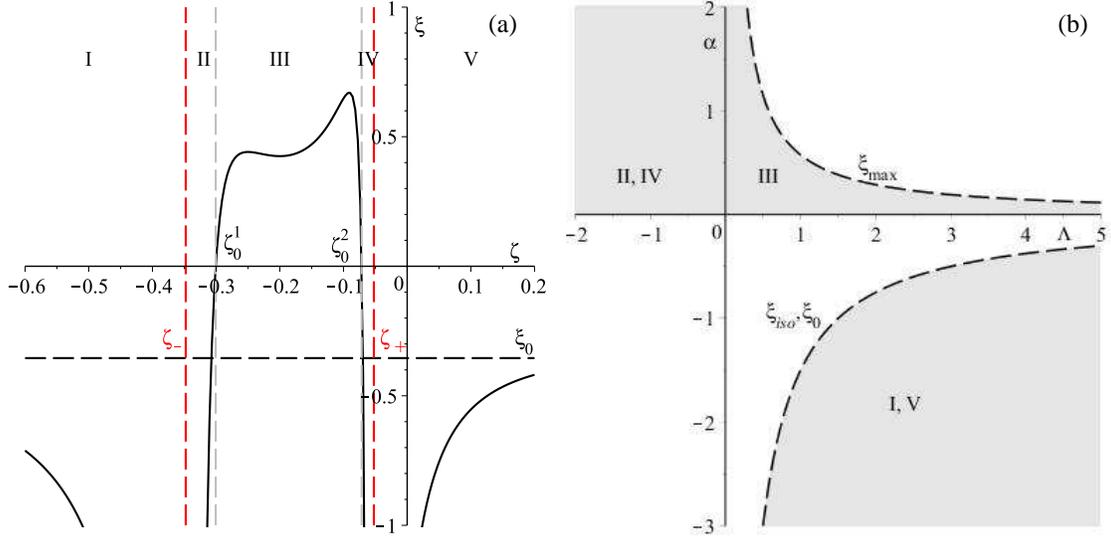}
\caption{Structure of the solutions for the general case on (a) panel and distribution of all possible $\{\alpha, \Lambda\}$ for stable solutions
(see the text for more details).}\label{fig5}
\end{figure}

The reported scheme and the results are illustrated in Fig.~\ref{fig5}. There on panel (a) we presented the structure of solutions on $\xi(\zeta)$ graph: vertical red dashed line represent vertical
asymptotes coming from zeros of $P_2$ (Eq. (\ref{eom2_theta})) and located at $\zeta_\pm$ (Eq. (\ref{eom2_denum_roots})); $\zeta_0^{1, 2}$ are zeros of $P_1$ (Eq. (\ref{eom2_xi})), their locations are
highlighted by grey dashed vertical lines; horizontal dashed black line corresponds to $\xi_0$ -- asymptotic value for $\xi(\zeta)$ at $\zeta\to\pm\infty$ (see Eq. (\ref{eom2_xi0})).

Then, according to the scheme, conditions for existence of the solutions could be described as follows (regions are numbered according to Fig.~\ref{fig5}(a)):

\begin{enumerate}

\item[I] for $\zeta < \zeta_-$ we have $\theta < 0$ (see Fig.~\ref{fig1}(b)) so that $\alpha < 0$, and $\xi < \xi_0$, so that solutions exist for $\alpha < 0$, $\Lambda > \alpha/\xi_0$;

\item[II] for $\zeta_- < \zeta < \zeta_0^1$ we have $\theta > 0$ (see Fig.~\ref{fig1}(b)) so that $\alpha > 0$, and $\xi < 0$, so that solutions exist for $\alpha > 0$, $\Lambda < 0$;

\item[III] for $\zeta_0^1 < \zeta < \zeta_0^2$ we have $\theta > 0$ (see Fig.~\ref{fig1}(b)) so that $\alpha > 0$, and $0 < \xi < \xi_{max}$, so that solutions exist for $\alpha > 0$,
$0 < \Lambda < \alpha/\xi_{max}$;

\item[IV] for $\zeta_0^2 < \zeta < \zeta_+$ we have $\theta > 0$ (see Fig.~\ref{fig1}(b)) so that $\alpha > 0$, and $\xi < 0$, so that solutions exist for $\alpha > 0$, $\Lambda < 0$;

\item[V] finally, for $\zeta > \zeta_+$ we have $\theta < 0$ (see Fig.~\ref{fig1}(b)) so that $\alpha < 0$, and $\xi < \xi_{iso}$, so that solutions exist for $\alpha < 0$, $\Lambda > \alpha/\xi_{iso}$
(unlike region I, in region V maximal value for $\xi$ in bounded by $\xi_{iso}$, see Fig.~\ref{fig1}(d) and Fig.~\ref{fig6}).

\end{enumerate}

Existence regions are collected in Fig.~\ref{fig5}(b). Regions I and V are unbound on $\zeta$, as well as on $\alpha$ and $\Lambda$ from below
(though there is a upper limit on $\alpha\Lambda$). Regions I, III
and V require $\Lambda > 0$, only II and IV exist for $\Lambda < 0$. From Fig.~\ref{fig5}(a) one can see that regions II and IV exist for a very narrow ranges of $\zeta < 0$, but these ranges cover entire
$\Lambda < 0$. So that there could exist two solutions in II and IV regions (one from II and one from IV), two solutions in I and V regions and up to four solutions in III region
(they do not sum up, though---say, solutions in III region exist for $\xi > 0$ while in other regions---for $\xi < 0$, see Fig.~\ref{fig5}(a)).
Depending on $\xi$,
for $\xi < \xi_0$ there are four solutions (one from each of I, II, IV and V regions); for $\xi_0 < \xi < \xi_{iso}$ there are again four solutions but now one from II and IV plus two from V regions;
for $\xi_{iso} < \xi < 0$ there are two solutions in II and IV regions, for $0 < \xi < \xi(\zeta_{min})$ there are also two
solutions (both from III region) and for $\xi(\zeta_{min}) < \xi < \xi_{max}$ there could be up to four solutions; the number of solutions is obtained using $\xi = \const$ lines in Fig.~\ref{fig5}(a).

Now let us discuss how this scheme is affected by additional stability requirement. As we derived earlier, for the solution to be stable, we require either $\zeta > \zeta_2$, $h > 0$ or
$\zeta < \zeta_2$, $h < 0$, where $\zeta_2 = -q/p$ and depending on $\{p, q\}$ it is located at either local maximum, local minimum or global maximum within region III (see Section~\ref{exist} for
details). So that the described above scheme is altered a bit: now we do not choose sign for $|h|$ but it is determined from stability requirement: if $\zeta > -q/p$ then $h>0$, if $\zeta < -q/p$
then $h<0$; the rest of the scheme remains the same.

The abundance of the solutions is halved, though: for $h>0$ only solutions located to the right of the separatrix line are stable while for $h<0$ -- only those to the left. So that for each case
we might have one solution from I (for $h<0$) or up to two from V (for $h>0$) regions, one solution from II (for $h<0$) or IV (for $h>0$) regions
and up to three solutions within III region, depending on
$p$, $q$ and sign of $h$. The conditions for stable solutions are barely affected: since we anyway have solutions from I or V and II or IV regions, 2nd quadrant of Fig.~\ref{fig5}(b)
remains the same while for 4th quadrant we choose $\xi_0$ for region I ($h<0$) and $\xi_{iso}$ for region V ($h>0$) (see also Fig.~\ref{fig6}).
Similar situation is with III region (1st quadrant of Fig.~\ref{fig5}(b))---there $\xi_{max}$ for $h<0$ is replaced with another (yet still positive) value according to the scheme
described in Section~\ref{sec_stab}; for $h>0$ it remains $\xi_{max}$.

Concluding, for the general $\{p, q\} > 2$ case we always have stable exponential solutions for $\alpha > 0$, $\Lambda < \xi_{max}/\alpha$ as described above (including entire $\Lambda < 0$),
and for $\alpha < 0$, $\Lambda > \xi_0/\alpha$ with $\xi_0$ defined in (\ref{eom2_xi0}).

Still there remain cases with $\{p, q\} = \{1, 2\}$, described in Section~\ref{sec_spec} For $p=1$, $q=1$ we have just 3D gravity which is ill-defined within EGB gravity. For $q=1$ and arbitrary $p$
we have unphysical situation: the obtained solution is valid for any $h$, which means that it is physically cannot be reached, as discussed in~\cite{my16b}. So that the cases when one of the subspaces is
one-dimensional, are pathological.

For the $p=2$, $q=2$ case we have solutions only for $\Lambda > 0$ (though, both signs for $\alpha$ are allowed). Finally, for $q=2$ and arbitrary $p>2$ (so that one of subspaces is two-dimensional)
we have situation similar to the described scheme but with unbounded $\xi$ within $\zeta < 0$ (see Fig.~\ref{fig3}).

Finally, the last case which worth separate mentioning is the case with $q=3$ -- this case could represent realistic compactification---indeed, three-dimensional subspace would represent ``our Universe''
and it should expand while $p$-dimensional subspace represent extra dimensions and should contract. This situation is explored in Section~\ref{sec_3D}; as $q=3 > 2$ formally it falls within general scheme,
but as it has direct physical meaning we decided to report it separately. Also there we have fixed $h > 0$ (three-dimensional subspace -- ``our Universe'' -- is expanding) and $\zeta < 0$ (so that to have
$H < 0$: extra dimensions should contract---this explains why we do not see them). This leaves us with a narrow range for $\zeta$: $\zeta > - p/3$ (for the solution to be stable) and $\zeta < 0$ (to
have physical meaning); detailed situation is presented in Fig~\ref{fig4}(a). Still, this region on $\zeta$ is split into three subregions, each corresponding to different
sign combinations of $\alpha$ and $\Lambda$, presented in Fig~\ref{fig4}(b). One can see that it resembles the generic $\{p, q\} > 2$ case, presented in Fig~\ref{fig5}, but, again, since
this specific case particularly physically meaningful, we reported it separately.

This concludes our study of the exponential solutions in the setup with two subspaces. This particular case with two subspaces is extremely important in lower number of
spatial dimensions (lower than around seven) since for this case there are no anisotropic exponential solutions of other sorts. In higher number of dimensions, one could build exponential solutions
with three and higher number of independent subspaces, but so far it is an open question under which conditions initially totally anisotropic (Bianchi-I-type) Universe ends up in different
anisotropic exponential solutions; study of this question was initiated in~\cite{PT2017} (see also~\cite{CGT-2020}),
but it was done in lower number of dimensions, where only solutions with two subspaces exist. Nevertheless, even in
higher dimensions solutions with two subspaces still exist---they are complimented by solutions with more subspaces---and since they also exist there, it is important to know their properties,
conditions for stability and so on, which is done in this paper. In the papers to follow we are going to consider more complicated cases---with three and more subspaces with independent evolution.

{\bf Note added:} after this manuscript was submitted to arXiv, we became aware of previously published paper~\cite{IK}, where same problem was considered.
The case with $p=3$ was considered 
in~\cite{stab_add1} while $q=2$ case in~\cite{stab_add2}. We would like to thank V.D. Ivashchuk for bringing our attention to these publications.

\end{document}